\documentclass[12pt,preprint]{aastex}

\newcommand{\sollum} {\hbox{$L_{\odot}$}}
\newcommand{\solmass} {\hbox{$M_{\odot}$}}
\def\deg{\ifmmode^\circ\else$^\circ$\fi}

\shorttitle{Discovery of Eight Extragalactic H$_2$O Masers}
\shortauthors{Braatz \& Gugliucci}

\begin{document}

\title{The Discovery of Water Maser Emission from Eight Nearby Galaxies}


\author{J.A. Braatz}
\affil{National Radio Astronomy Observatory, 520 Edgemont Rd., Charlottesville, VA 22903}
\email{jbraatz@nrao.edu}

\and

\author{N.E. Gugliucci}
\affil{Department of Astronomy, University of Virginia, Charlottesville, VA 22904}
\email{neg9j@virginia.edu}

\begin{abstract}
Using the Green Bank Telescope, we conducted a ``snapshot'' survey for water
maser emission toward the nuclei of 611 galaxies and detected eight new
sources.  The sample consisted of nearby ($v < $ 5000 km s$^{-1}$) and luminous
($M_B$ $<$ -19.5) galaxies, some with known nuclear activity but most
not previously known to host AGNs.  Our detections include both 
megamasers associated with AGNs and relatively low luminosity masers
probably associated with star formation.
The detection in UGC~3789 is particularly intriguing
because the spectrum shows both systemic and high-velocity lines indicative
of emission from an AGN accretion disk seen edge-on.  Based on six months of
monitoring, we detected accelerations among the systemic features ranging 
from 2 to 8 km s$^{-1}$ yr$^{-1}$, the larger values belonging to the
most redshifted systemic components.  High-velocity maser lines
in UGC~3789 show no detectable drift over the same period.  Although
UGC~3789 was not known to be an AGN prior to this survey, the presence of a
disk maser is strong evidence for nuclear activity, and an optical spectrum
obtained later has confirmed it.  With follow up observations, it may be 
possible to measure a geometric distance to UGC~3789.
\end{abstract}

\keywords{galaxies: active --- galaxies: nuclei --- galaxies: Seyfert ---
ISM: molecules --- masers --- radio lines: galaxies}

\section{Introduction}

Water maser emission at 22 GHz is commonly detected toward Galactic H II
regions, pinpointing the locations of massive star formation.  The median
isotropic line luminosity in these sources is 10$^{-4}$ $\sollum$ (Genzel \& Downes, 1977), although in
exceptional cases such as W49 the luminosity can reach $\sim$ 1 $\sollum$.
Analogs to the W49 maser have been observed in H II regions of several
nearby galaxies as well, where the combined radiation from multiple sites
of star formation can lead to somewhat larger luminosities.  The most distant (14.5 Mpc)
and luminous (L$_{H_2O}$ $\sim$ 8 $\sollum$) example of this type was detected
toward the starburst galaxy NGC~2146 (Tarchi et al. 2002).  In recent years,
masers in extragalactic star-forming regions have become an important tool for
measuring proper motions of the host galaxies, as in IC~10 and M~33
(Brunthaler et al. 2005, 2007)

Extragalactic water masers are also detected toward some active galactic
nuclei (AGN), usually Seyfert 2 or LINER galaxies (Braatz et al. 1997).
These sources have been called ``megamasers'' because the isotropic line 
luminosities exceed the median value of Galactic masers by $\sim$ 6 orders 
of magnitude.  Both galactic and extragalactic masers, however, are not 
believed to emit isotropically so the true line luminosity is smaller in each 
case by an unknown beaming factor.  Masers in AGN accretion disks, 
specifically, are thought to be beamed in the plane of the disk.

High resolution imaging shows that megamasers are located within parsecs of the
galactic nucleus, where they are associated either with jet-cloud
interactions, nuclear outflows, or parsec-scale accretion disks.  
Unlike star formation masers, which only have Doppler components detected within $\sim$
200 km s$^{-1}$ of the galaxy systemic recession velocity, masers in AGN
accretion disks have shown velocity offsets up to $\sim$ 1300 km s$^{-1}$ 
(Kondratko et al. 2006a) owing to rotation about the central
black hole.  In the case of NGC~4258, masers trace the accretion disk 
0.11 -- 0.28 pc from the black hole (Argon et al. 2007), and have been
used to measure the most precise geometric distance to a galaxy,
7.2 $\pm$ 0.2 (random statistical) $\pm$ 0.5 (systematic) Mpc (Herrnstein et al. 1999).
The topic of extragalactic masers has recently been reviewed by Lo (2005).

NGC~4258 is currently the only galaxy with a robust maser distance, but large surveys
are seeking additional examples of disk masers appropriate for such a measurement.
Most surveys target Seyfert 2 and LINER galaxies (e.g. Greenhill et al. 2003; Braatz et al. 2004; 
Kondratko et al. 2006a, 2006b).  With sensitive
observations using large aperture telescopes, overall detection rates
are $\sim$ 5\%, and several disk masers have been identified.  The masers detected
by these surveys tend to be faint (peak fluxes of a few tens of mJy) and challenging 
to study in detail.  Candidates for new surveys are limited by the number of AGNs 
identified in optical surveys.

While disk masers are clearly associated with active galaxies, the luminosity of the host AGN
is modest in several of the most interesting cases.  For example, the bolometric
luminosity of NGC~4258, which has been classified either as a LINER or a weakly active 
Sy 1.9, is only $10^{42}$ erg s$^{-1}$ (e.g. Chary et al. 2000).  Furthermore, Ho et al.
(1997) found that 43\% of a magnitude-limited sample of nearby galaxies 
host AGN, mostly with low luminosity, demonstrating that many apparently normal galaxies in fact
host low luminosity AGN.  We were thus motivated to conduct a new survey for H$_2$O maser
emission in nearby galaxies not necessarily known to host AGNs.  Our intention was to identify
bright maser systems ($\gtrsim$ 50 mJy) conducive to detailed Very Long Baseline Interferometry 
(VLBI) imaging and spectral line monitoring.  With the GBT we can identify such masers using
short integrations, making it possible to observe a large sample and balance the relatively
low detection rate expected.

\section{Observations}
We conducted observations between 2005 September 27 and 2007 April 14 
using the Green Bank Telescope (GBT) of the National Radio Astronomy 
Observatory.  The project was observed in 31 individual sessions,
often assigned to fill short gaps in the telescope 
schedule.  We configured the spectrometer with two 200 MHz spectral windows, 
each with 8192 channels, giving a channel spacing of 24 kHz corresponding to
0.33 km s$^{-1}$ at 22 GHz.  The spectral windows were overlapped slightly such that the
total bandwidth coverage was 380 MHz (5100 km s$^{-1}$).  The systemic 
recession velocity of each galaxy was typically centered in one spectral 
window and the second window was offset to the red.  For a few galaxies, the 
spectral windows were tuned with the systemic recession velocity in the 
overlap region, giving symmetric velocity coverage.  We observed in dual
circular polarization using a total power nodding mode in which the galaxy was 
placed alternately in one of two beams of the 18 -- 22 GHz K-band receiver.  
A nodding cycle of 2.5 minutes per beam provided 5 minutes of integration per source.  
When time permitted, we obtained longer integrations to investigate possible detections.

We reduced and analyzed the data using GBTIDL.  In order to enhance weak 
signals, we smoothed the blank sky reference spectra using a 16-channel boxcar function
prior to calibration.  We corrected the spectra for atmospheric 
attenuation using opacities derived from weather data, and we applied an 
elevation-dependent gain curve.  The flux density scale uncertainty is about
15\%, limited by noise tube calibration and pointing errors.
To remove residual baseline shapes from each 200 MHz spectrum, 
we subtracted a polynomial (typically third order) fit to the line-free channels 
of the calibrated spectrum.  Typical rms noise levels for the 5-minute 
observations were 6 mJy per 0.33 km s$^{-1}$ channel, after Hanning smoothing.

Our candidate galaxies were selected from the Third Reference Catalog of Bright Galaxies 
(de Vaucouleurs et al. 1991) using coarse criteria intended primarily to eliminate 
galaxies without well-defined nuclei and those unlikely to host a hidden AGN.
The sample has (1) systemic recession velocity $v <$ 5000 km s$^{-1}$, which
corresponds to the distance at which the GBT could detect the NGC~4258 maser in
5 minutes; (2) Hubble stage --3 $<$ T $<$ +5 to choose lenticulars and early-type
spirals; (3) absolute magnitude M$_B$ $<$ --19.5 to eliminate dwarf galaxies and 
boost the likelihood of a prominent nuclear black hole; and (4) declinations north 
of --30$\deg$, accessible to the GBT.  We also eliminated galaxies previously searched 
with the GBT.  These criteria identified 1074 candidate galaxies, not all of which 
were observed in the allotted time.  We plan to observe the remaining galaxies in 
ongoing surveys.  Galaxies were finally chosen for observation based on nearest 
neighbors, to minimize telescope slew time.

\section{Results}

We searched for 22 GHz H$_2$O maser emission toward the nuclei of 611 galaxies
and detected masers in eight.  Table 1 lists properties of the detected masers and
Table 2 lists the undetected galaxies from the survey, along with relevant observing parameters.  
Each new maser was detected in more than one observing session except the one in NGC~1106, 
which was not identified until the data were reexamined at the conclusion of the scheduled 
observing.  The best spectrum for each of the newly detected maser systems is shown in Figure 1.
Velocities quoted in this paper are heliocentric and use the optical definition of Doppler shift.

Comments on the new masers and their host galaxies follow.  For several of the
galaxies, we discuss the maser profile in the context of an accretion disk maser.  In this 
model, ``systemic'' emission refers to masers on the near side of the edge-on disk, along 
the line of sight to the black hole.  ``High-velocity'' emission refers to masers located 
on the midline of the disk, rotating directly toward or away from the observer.  

\subsection {NGC 23}

The maser detected in NGC~23 exhibits a broad profile relatively free of isolated, narrow features
and centered near the systemic velocity (Figure 1).  The spectrum shows no evidence of high-velocity 
lines.  The isotropic luminosity of the source is 180 $\sollum$, suggesting the maser is
associated with an AGN.  The profile resembles those from masers associated with nuclear
jets, such as NGC~1052 (Claussen et al. 1998) and Mrk~348 (Peck et al. 2003).  However, NGC~23 
is not known to have an AGN.  With 
log $L_{IR}$ = 11.05 $\sollum$ (Sanders et al. 2003), NGC~23 is a luminous infrared galaxy 
(LIRG).  The intense star formation associated with LIRGs might be expected to produce a large 
number of star-formation masers whose cumulative emission could reach the observed isotropic 
luminosity in NGC~23.  However, Zhang et al. (2006) point out that the ultraluminous infrared 
galaxy (ULIRG) Arp 220 would have been detected easily in recent surveys if the maser emission 
were correlated with IR luminosity, yet Arp 220 has not been detected in H$_2$O.
Two ULIRGs have been detected in the water maser line: UGC~5101 (Zhang et al. 2006) 
and NGC~6240 (Hagiwara et al. 2002; Nakai et al. 2002; Braatz et al. 2003).  In both cases
the masers appear to be associated with an AGN and not star formation (Hagiwara et al. 2003;
Zhang et al. 2006).  High resolution imaging could resolve the issue of whether 
the maser in NGC~23 is associated with star formation or an AGN.

\subsection {NGC 1106}

NGC~1106 is an emission line galaxy, called a Seyfert 2 by Bonatto et al. (1996).
The maser detected in this galaxy contains a single, narrow (1.0 km s$^{-1}$ FWHM) component at 
3814 km s$^{-1}$, offset by 523 km s$^{-1}$ from the 4337 km s$^{-1}$ systemic recession velocity.
The large velocity offset hints that the maser may be a high-velocity line in an accretion disk.
The peak flux density is 74 mJy.  Because the maser was not identified until all observations had 
been completed, we did not obtain a confirmation spectrum for this source.  However, the
maser line appears in both polarizations and in both beams of our observation.  An observation of 
NGC~1106 by Kondratko et al. (2006b) with 13 mJy rms sensitivity failed to detect the line, but it 
would have appeared only at about 4$\sigma$ given their channel spacing of 1.3 km s$^{-1}$.  We
also note that variability is common in narrow high-velocity lines.

\subsection {UGC 3193}

The maser observed toward the barred spiral galaxy UGC~3193 spans $\sim$ 350 km s$^{-1}$
and is localized in four distinct velocity windows.  The profile shows remarkable symmetry 
about the galaxy's recession velocity, even in the details of the inner two complexes.  
However, no lines are detected at the systemic velocity itself.  The profile thus hints that 
the maser originates in a slowly rotating, edge-on disk, with only high-velocity features detected.  
A spiral density pattern in an accretion disk or discrete rings could produce a pattern
as observed.  The absence of emission directly at the systemic velocity may indicate that the 
disk inclination or a warp prevents beaming of masers on the near side of the disk into our 
line of sight.  Alternatively, a lack of background continuum emission that would otherwise 
provide seed emission for systemic features could account for the absence of detectable maser 
emission near the systemic velocity.

\subsection {UGC 3789}

The maser detected in UGC~3789 shows a profile characteristic of emission from an edge-on nuclear 
accretion disk.  Maser lines are detected in three distinct velocity ranges (Figure 1), roughly 
symmetric about the galaxy's recession velocity of 3325 $\pm$ 24 km s$^{-1}$.  Systemic lines are 
centered near 3275 km s$^{-1}$, offset slightly from the reported recession velocity.  High velocity 
lines span $\sim$ 1500 km s$^{-1}$, suggesting the maser disk traces rotation speeds 
up to $\sim$ 750 km s$^{-1}$.

Upon associating the maser in UGC~3789 with disk emission, we initiated a program to observe the
source at roughly 3-week intervals in order to track the velocities of systemic maser
components and measure the centripetal acceleration of the maser gas in orbit around the central
black hole.  The
initial results of this monitoring program are shown in Figures 2 and 3.  With the 
six months of data presented here we have measured accelerations for nine individual lines 
among the systemic features.  The accelerations, listed in the Figure 2 caption, range 
from 1.8 -- 8.1 km s$^{-1}$ yr$^{-1}$, with the smallest accelerations measured 
near $\sim$ 3245 km s$^{-1}$ on the blue edge of the systemic features and the largest 
near $\sim$ 3310 km s$^{-1}$ on the red edge.  The mean of the nine accelerations 
identified in Figure 2 is 4.0 $\pm$ 0.8 km s$^{-1}$ yr$^{-1}$.  The wide range of measured 
accelerations likely indicates that the systemic masers are not confined to a radially thin 
ring, but rather are spread over a range of radial distance from the dynamical center of 
the galaxy.  The acceleration of systemic masers in NGC~4258 also varies for different
components, although with a smaller range (e.g. Humphreys et al. 2008).  Among the 
high-velocity features, measured accelerations are all less than 0.3 km s$^{-1}$ yr$^{-1}$.  
Additional monitoring observations of UGC~3789 and a more detailed analysis of the accelerations will 
follow in a later paper.

A robust measurement of the mass of a black hole can be made by measuring the rotation 
curve of a maser disk from a VLBI map.  Without VLBI we can still make an estimate.
If we suppose that the systemic masers with the largest measured acceleration occur at 
the same orbital radius as gas producing the largest rotation velocities among the 
high-velocity masers and that the accretion disk is edge-on, we can approximate $M_{bh}$, 
the mass of the black hole, by
$M_{bh} = rv^2/G = v^4/aG$.  Here $r$ is the radius of the orbiting gas, $v$ is its 
rotation velocity, $a$ is the centripetal acceleration, and $G$ is the gravitational constant.
Then $v \sim$ 750 km s$^{-1}$ and $a \sim$ 8.1 km s$^{-1}$ yr$^{-1}$ gives $M_{bh} \sim$ 
9 $\times$ 10$^6 \solmass$ for UGC~3789.

The systemic maser features in UGC~3789 are themselves split at $\sim$ 3275 km s$^{-1}$ 
by a dip in the emission (Figure 3).  
At times, the systemic emission in NGC~4258 has shown a similar dip (e.g. Greenhill 
et al. 1995a), although later data showed no evidence for it in that 
galaxy (Bragg et al. 2000).  Watson \& Wallin (1994) and 
Maoz \& McKee (1998) developed models that invoke an absorbing layer of noninverted 
H$_2$O in the maser disk to account for a dip in the emission.

Megamaser emission is a beacon of nuclear activity.  While UGC~3789 had not 
been identified as an AGN prior to our survey, an optical spectrum of the nucleus 
of the galaxy (Macri, private communication) reveals line flux ratios, 
such as [O III]5007/H$\beta$ and [O I]6300/H$\alpha$, and line widths of
H$\alpha$ and H$\beta$ indicative of a Seyfert 2 nucleus.

With additional observations, it might be possible to measure the distance to UGC~3789
using the maser method applied to NGC~4258 (Herrnstein et al. 1999).
We are continuing GBT monitoring and collecting sensitive VLBI observations for this
purpose, and the results of these studies will be presented in later papers.

\subsection {NGC 2989}

The weak maser emission detected in NGC~2989 spans 30 km s$^{-1}$ and is centered near the
systemic recession velocity of the galaxy.  No high velocity lines are detected.
Although the isotropic luminosity (40 $\sollum$) is large compared to masers in
star formation regions, this galaxy has a nuclear H II region (Veron-Cetty \& Veron, 2003) and the
line profile is otherwise consistent with maser emission associated with star formation.
If confirmed as a star formation maser, this galaxy, at a nominal distance of 58 Mpc 
(assuming H$_0$ = 72 km s$^{-1}$ Mpc$^{-1}$) would be the most distant known source of such 
emission.  However we cannot rule out the possibility that the maser is associated with an AGN.

\subsection {NGC 3359}

The maser emission from NGC~3359 has several weak, narrow features spanning $\sim$ 10 km s$^{-1}$ 
with a total isotropic luminosity of 0.7 $\sollum$, comparable to the luminosity of the strongest 
masers from H II regions in the Milky Way.  No high-velocity lines are evident in the maser
spectrum.  NGC~3359 is an H II galaxy, and the maser is consistent with star formation.

\subsection {NGC 4527}

In addition to containing a LINER (Ho et al. 1997), the galaxy NGC~4527 is bright
in the far infrared (S$_{100 \mu m}$ = 64 Jy; Fullmer \& Lonsdale 1989).  Such far-infrared emission is 
commonly associated with dust heated by young, massive stars.  The galaxy has been searched for 
H$_2$O maser emission previously by Kondratko et al. (2006b) and Henkel et al. (2005) but went 
undetected.  The rms sensitivities in their observations were
12 mJy per 1.3 km s$^{-1}$ channel and 15 mJy per 1.1 km s$^{-1}$ channel,
respectively.  The modest isotropic luminosity (4 $\sollum$) is 
consistent with maser emission from a star formation region, but the velocity offset from the 
systemic recession velocity (170 km s$^{-1}$) leaves open the possibility that the maser is 
associated with the LINER.  We note that NGC~4527 was host to the peculiar type Ia supernove 1991T.

\subsection {NGC 7479}

With a type 1.9 Seyfert nucleus (Ho et al. 1997), NGC~7479 has
been targeted in previous surveys for H$_2$O masers in AGNs, including those by Kondratko et al. 
(2006b) and Braatz et al. (1996), in which the rms sensitivities were 19 mJy per
1.3 km s$^{-1}$ channel and 54 mJy per 0.66 km s$^{-1}$ channel, respectively.  
We detected the 140 mJy line at 2532 km s$^{-1}$ in the initial 5-minute integration, 
and the deeper spectrum shown in Figure 1 then revealed a number of distinct, weaker components
redward of the systemic velocity.  If the lines originate 
in an edge-on nuclear disk, the implied rotation speeds range up to 280 km s$^{-1}$.

\section{Discussion}

Megamasers in accretion disks can play an important role in
cosmology.  As a complement to observations of the cosmic 
microwave background, a precise ($\lesssim$ 3\%) determination of 
the Hubble constant would provide a strong constraint on several 
key cosmological parameters, including the equation of state of dark 
energy (Hu 2005; Spergel et al. 2007).  Measuring distances to 
water maser galaxies can contribute to improving the Hubble constant 
in two ways.  First, maser distances to nearby galaxies (D $\lesssim$ 30 Mpc) 
such as NGC~4258 can be used to refine the absolute calibration of 
Cepheid variables and replace or supplement the Large Magellanic
Cloud as the primary anchor in the extragalactic distance scale (Macri et al. 2006).  
Second, maser distances to galaxies in the Hubble flow can be used to
measure H$_0$ directly.  Because maser distances are not based on standard
candles and individual uncertainties should be uncorrelated, it may be
possible to reduce the total error in H$_0$ by measuring many such
distances (e.g. Greenhill 2004).  So far, however, a maser distance has
only been possible for NGC~4258 (Herrnstein et al. 1999), and it is too
close to measure H$_0$ directly.

Measuring maser distances to galaxies requires a robust model
of the rotation structure and geometry of the accretion disk from VLBI 
imaging and either proper motions of maser features on the near side of the 
disk, or centripetal accelerations measured by spectral line monitoring.
While proper motions ($\sim$ 30 $\mu$as yr$^{-1}$ in NGC~4258 at 7.2 Mpc,
Herrnstein et al. 1999) might be challenging for galaxies in the Hubble 
flow, measuring accelerations is possible, as demonstrated here for UGC~3789.
We note, however, that in UGC~3789 additional monitoring will be necessary 
to properly model the radial structure implied by the variation in 
accelerations among different maser components.

Prospects for mapping and modeling the disk in UGC~3789 based on VLBI
observations can only be assessed when that imaging is available.
However, it is clear from the large number of bright ($\gtrsim$ 20 mJy) 
maser features within each of the three spectral complexes that the rotation 
curve can be well sampled.  Moreover, the mean acceleration ($\sim$ 
4.0 km s$^{-1}$ yr$^{-1}$) and mean rotation velocity ($\sim$ 600 km s$^{-1}$)
suggest a disk radius of $\sim$ 0.09 pc.  At a nominal distance
to UGC~3789 of 46 Mpc, this corresponds to $\sim$ 0.4 mas. Such a
maser disk can be resolved by VLBI at 22 GHz.

\section {Summary}

In this survey for extragalactic water masers we employed a ``snapshot'' strategy 
designed to detect bright emission from nearby galaxies that may host hidden nuclear 
activity.  Although the detection rate was low (1.3\%) we observed a large
number of galaxies and detected eight new masers.  While it is not possible
to assess the nature of these systems until they are imaged with interferometers, 
several appear particularly interesting.  If confirmed as a star formation maser,
the source in NGC~2989 would be the most distant and luminous example of that class.
The maser in NGC~23 has a profile that resembles those of jet masers, but it is 
a LIRG and not known to host an AGN.  In UGC~3193 the maser profile is remarkably
symmetric but shows no emission directly at the systemic velocity of the galaxy.
The most intriguing detection is in UGC~3789, which was not previously known to
be an AGN.  Its spectral profile is characteristic of maser emission from an accretion disk,
and we have made a preliminary measurement of accelerations for several of its 
systemic features.  With future VLBI studies and disk modeling, it may be possible to measure 
a geometric distance to this galaxy based on observations of its maser disk.

\acknowledgements
We thank Lucas Macri for providing an optical spectrum of UGC~3789.
We are grateful for the many contributions from the NRAO staff in Green Bank.
We also thank Jim Condon, Lincoln Greenhill, Mark Reid, Fred Lo, and Christian
Henkel for useful discussions.  
The National Radio Astronomy Observatory is a facility of the National 
Science Foundation operated under cooperative agreement by Associated
Universities, Inc.
This research has made
use of the NASA/IPAC Extragalactic Database (NED) which is operated by the Jet
Propulsion Laboratory, California Institute of Technology, under contract with
the National Aeronautics and Space Administration.

\clearpage

\begin{table}
\caption{H$_2$O Masers Detected During the GBT Survey}
\bigskip
\begin{tabular}{lrrrrrr}
\tableline\tableline
Source & $\alpha_{2000}$ & $\delta_{2000}$ & V$_{hel}$\tablenotemark{a} & $S_{peak}$\tablenotemark{b} & L$_{iso}$\tablenotemark{c} & Date of Observation \\
 & (\ h\ \ \ m\ \ \ s\ )&( $\ \deg\ \ \ \arcmin\ \ \ \arcsec\ \ $) & (km s$^{-1})$ & (mJy) & ($\sollum$) &  \\
\tableline
NGC~23   & 00 09 53.4 & +25 55 26 &  4566 $\pm$ 2  & 51  & 180 & 26 Nov 2006 \\
NGC~1106 & 02 50 40.5 & +41 40 17 &  4337 $\pm$ 19 & 74  & 8   & 20 Nov 2005 \\
UGC~3193 & 04 52 52.6 & +03 03 26 &  4454 $\pm$ 10 & 130 & 270 & 20 Nov 2005 \\
UGC~3789 & 07 19 30.9 & +59 21 18 &  3325 $\pm$ 24 & 134 & 370 & 21 Dec 2005 \\
NGC~2989 & 09 45 25.2 & -18 22 26 &  4146 $\pm$ 5  & 36  & 40  & 20 Nov 2005 \\
NGC~3359 & 10 46 36.8 & +63 13 25 &  1014 $\pm$ 1  & 40  & 0.7 & 31 Dec 2005 \\
NGC~4527 & 12 34 08.5 & +02 39 14 &  1736 $\pm$ 1  & 16  & 4   & 14 Apr 2007 \\
NGC~7479 & 23 04 56.7 & +12 19 22 &  2381 $\pm$ 1  & 138 & 19  & 30 Dec 2006 \\
\tableline
\end{tabular}
\tablenotetext{a}{Heliocentric recession velocity from the NASA/IPAC Extragalactic
Database}
\tablenotetext{b}{Peak flux density}
\tablenotetext{c}{Inferred isotropic luminosity of the maser emission.}
\end{table}

\clearpage

\begin{deluxetable}{lccclcccrcr}
\tabletypesize{\scriptsize} 
\tablewidth{0pt} 
\tablecolumns{11}
\tablecaption{Galaxies surveyed for water maser emission but not detected.}
\tablehead{\colhead{Source} & \colhead{$\alpha_{2000}$} & \colhead{$\delta_{2000}$} &
\colhead{Velocity} & \colhead{Date} & \colhead{T$_{\mbox{sys}}$} & \colhead{Time} & 
\colhead{rms$_1$} & \colhead{V$_1$ range} & \colhead{rms$_2$} & \colhead{V$_2$ range}
\\ \colhead{} & \colhead{(\ h\ \ \ m\ \ \ s\ )} & \colhead{( $\ \deg\ \ \ \arcmin\ \ \ \arcsec\ \ $)}
& \colhead{(km~s$^{-1}$)} & \colhead{} & \colhead{(K)} & \colhead{(min)} & \colhead{(mJy)} &
\colhead{(km~s$^{-1}$)} & \colhead{(mJy)} & \colhead{(km~s$^{-1}$)}
}
\startdata
       NGC~7805 &  0 01 26.76 & +31 26 01.4 & 4850 & 2006 Jan 5 & 43.8 &  5 &  6.0 &  3464 -- 6248 &  5.7 &  5968 -- 8799\\
       NGC~7806 &  0 01 30.06 & +31 26 30.7 & 4768 & 2006 Jan 5 & 42.2 &  5 &  5.7 &  3383 -- 6166 &  5.7 &  5885 -- 8715\\
       NGC~7817 &  0 03 58.91 & +20 45 08.3 & 2309 & 2005 Dec 14 & 42.4 &  5 &  6.5 &   946 -- 3684 &  7.2 &  3408 -- 6192\\
       NGC~7819 &  0 04 24.54 & +31 28 19.3 & 4958 & 2006 Jan 5 & 39.7 &  5 &  5.7 &  3571 -- 6357 &  5.3 &  6077 -- 8910\\
          NGC~1 &  0 07 15.83 & +27 42 29.1 & 4545 & 2005 Dec 14 & 41.1 &  5 &  6.6 &  3162 -- 5941 &  7.7 &  5661 -- 8486\\
         NGC~12 &  0 08 44.81 & +04 36 44.9 & 3940 & 2005 Oct 2 & 60.6 &  5 &  8.5 &  3802 -- 6592 &  9.0 &  1333 -- 4078\\
         NGC~13 &  0 08 47.73 & +33 26 00.0 & 4808 & 2006 Oct 30 & 41.4 &  5 &  5.7 &  3422 -- 6206 &  5.3 &  5926 -- 8755\\
          NGC~9 &  0 08 54.69 & +23 49 01.3 & 4528 & 2005 Dec 14 & 41.8 &  5 &  6.0 &  3145 -- 5923 &  5.9 &  5644 -- 8468\\
         NGC~16 &  0 09 04.28 & +27 43 45.9 & 3056 & 2005 Dec 14 & 42.2 &  5 &  5.7 &  1686 -- 4438 &  5.4 &  4161 -- 6958\\
         NGC~26 &  0 10 25.86 & +25 49 54.6 & 4589 & 2005 Dec 14 & 41.2 &  5 &  5.5 &  3205 -- 5985 &  5.5 &  5705 -- 8531\\
\enddata
\tablecomments{
Table 2 is published in its entirety in the electronic edition of 
the Astrophysical Journal.  A portion is shown here for guidance regarding its 
form and content.  Col (1): Source name. Cols. (2 -- 3): RA and Dec (J2000).
Col. (4): Recession velocity. Col. (5): Date of observation.  Col. (6): System 
temperature measured from the first spectral window.  Col. (7): Integration 
time. Col. (8): Rms noise measured for the first spectral window, after Hanning
smoothing.  Col. (9): Velocity range covered by the first window.  Col. (10): 
Rms noise measured for the second window, after Hanning smoothing.  Col. (11): 
Velocity range covered by the second window.  
}
\end{deluxetable}

\clearpage

\begin{figure}
\plotone{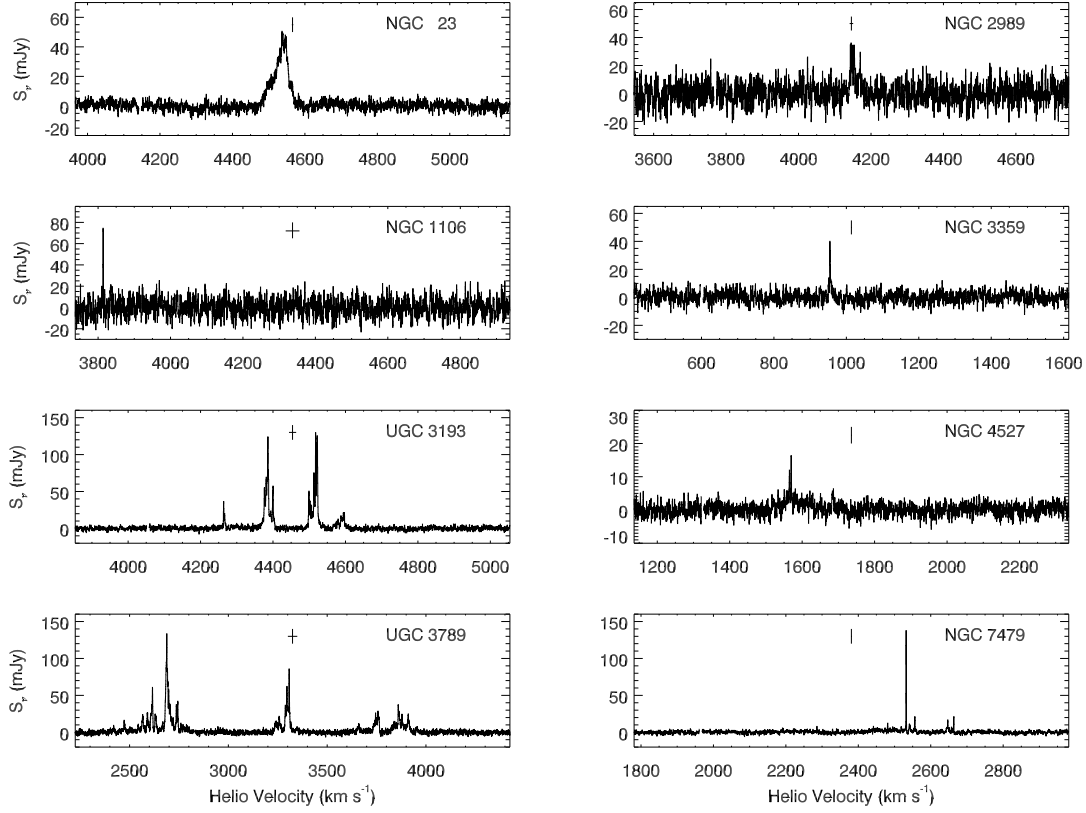}
\caption{Spectra of newly discovered 22 GHz water masers toward the nuclei of
8 galaxies.  The vertical line above each spectrum marks the systemic recession
velocity of the galaxy and its $\pm$ 1$\sigma$ uncertainty is indicated by the 
horizontal bar.  Each spectrum spans 1200 km s$^{-1}$ except
UGC~3789, which covers 2200 km s$^{-1}$.  Velocities are heliocentric and use
the optical definition of Doppler shift.
}
\end{figure}

\begin{figure}
\plotone{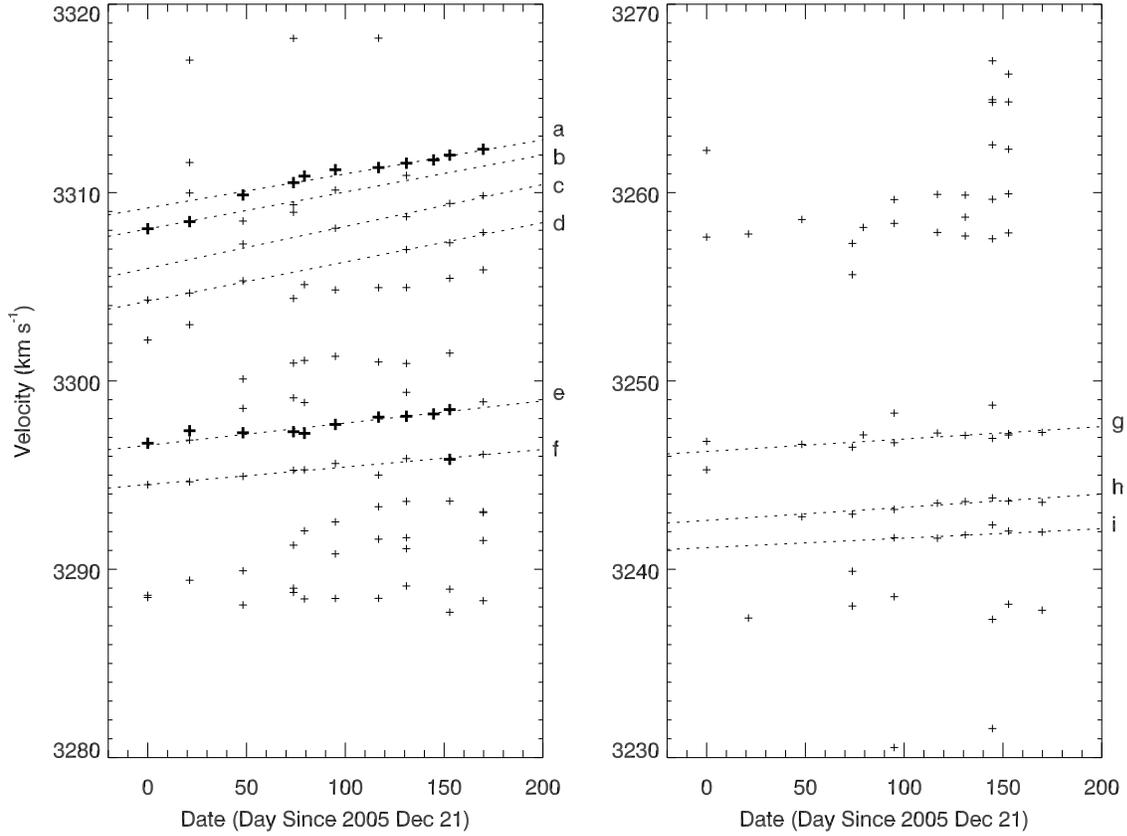}
\caption{The velocities of selected maser components in UGC~3789 are plotted as a 
function of date of observation, in days since the discovery date of 
2005 Dec 21.  The bold symbols denote prominent features with fluxes at 
least 30 mJy above emission at nearby frequencies.  Velocities
were determined by Gaussian fits.  Only narrow components distinguished from 
the base level of emission were fit, although in several cases we fit multiple
Gaussians to model blended features.  The plot is split into two panels for 
clarity.  The accelerations of the emission lines identified by dashed lines, 
measured by least squares linear fits and listed here in units of 
km s$^{-1}$ yr$^{-1}$, are 
(a) 6.6 $\pm$ 0.1; (b) 7.2 $\pm$ 0.2; (c) 8.1 $\pm$ 0.4; (d) 7.6 $\pm$ 0.2; 
(e) 4.3 $\pm$ 0.1; (f) 3.4 $\pm$ 0.1; (g) 2.4 $\pm$ 0.2; (h) 2.6 $\pm$ 0.3; and
(i) 1.8 $\pm$ 0.5.  The identification letters here correspond to the those
on the right side of each plot panel.}
\end{figure}

\begin{figure}
\plotone{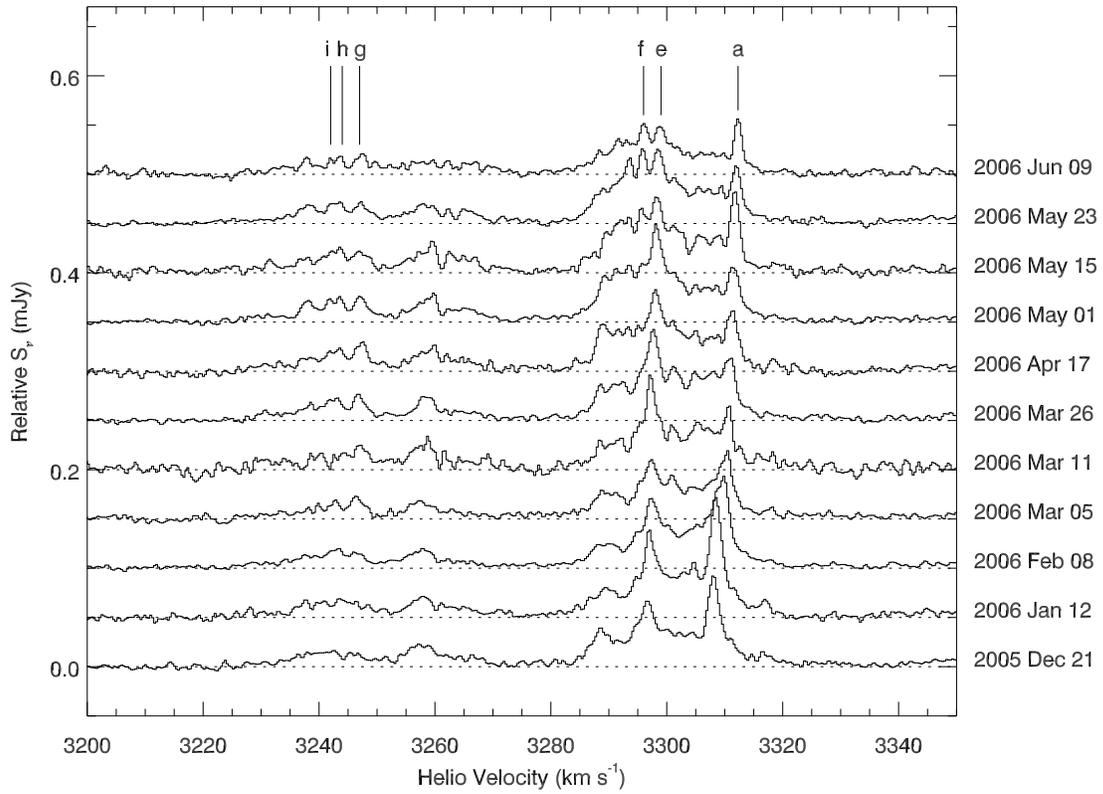}
\caption{Spectra of water maser emission from UGC~3789 near the systemic
recession velocity of the galaxy, spanning eleven epochs.  The spectra are
offset for clarity.  The labels at the top of the figure match individual features
to those identified in Figure 2.}
\end{figure}


\clearpage

\end{document}